\documentclass[prb,twocolumn,showpacs,superscriptaddress,preprintnumbers,amssymb]{revtex4}
\usepackage{graphicx}% Include figure files

\begin{document}
\title{Quantum Wells in Polar-Nonpolar Oxide Heterojunction Systems}
\author{C.-C Joseph Wang}
\affiliation{Department of Physics, The University of Texas at Austin, Austin, Texas, 78712-0264, USA}
\author{Bhagawan Sahu}
\affiliation{Microelectronics Research Center, The University of Texas at Austin, Austin Texas 78758, USA}
\author{Hongki Min}
\affiliation{Department of Physics, The University of Texas at Austin, Austin, Texas, 78712-0264, USA}
\author{Wei-Cheng Lee}
\affiliation{Department of Physics, The University of Texas at Austin, Austin, Texas, 78712-0264, USA}
\author{Allan H. MacDonald}
\affiliation{Department of Physics, The University of Texas at Austin, Austin, Texas, 78712-0264, USA}

\date{\today}
\begin{abstract}
We address the electronic structure of quantum wells in polar-nonpolar oxide heterojunction systems
focusing on the case of non-polar BaVO$_3$ wells surrounded by polar LaTiO$_3$ barriers.
Our discussion is based on a density functional
description using the local spin density approximation
with local correlation corrections (LSDA+U).  We conclude that a variety of quite different two-dimensional
electron systems can occur at interfaces between insulating materials depending on band line-ups and on
the geometrical arrangement of polarity discontinuities.
\end{abstract}

\maketitle

\section{Introduction}

Complex oxide materials exhibit a wide variety of exotic phenomena from
high temperature superconductivity\cite{HS}, to colossal magnetoresistance\cite{CM},
to electronic and orbital order\cite{MIT}.
Recently, inspired by advances in epitaxial growth techniques,
there has been increasing interest\cite{Huang,sc,Brinkman,Ahn,Spaldin,Pickett,WC} in making progress toward
the goal of controlling oxide properties by designing layered structures at the
atomic level.  The class of artificial materials which has been studied most
extensively is ABO$_3$ perovskites.  Already there is evidence for
the emergence of superconductivity from non-superconducting materials\cite{sc}
and magnetism from nonmagnetic materials\cite{Brinkman}.  One element which distinquishes
oxide heterojunction systems from weakly correlated semiconductor heterojunctions
is the possibility of exploiting gradients in polarity\cite{Sawatzky} to help control electronic properties.
In this article we study the electronic properties of oxide quantum wells in which the
barrier and well materials have very different polarities, using the case of
non-polar BaVO$_3$ wells surrounded by polar LaTiO$_3$ barriers as an example.

Our paper is organized as follows.
In Section II, we present a qualitative discussion of four possible polar-nonpolar
quantum well geometries.  We have tested conjectures made in Section II
by performing local spin-density-approximation
with local-correlation corrections (LSDA+U) calculations for BVO
quantum wells in LTO and LTO quantum wells in BVO.
Throughout this paper, we use LTO as shorthand for the perovskites associated with LaO and TiO$_{2}$ layers and BVO as shorthand
for the perovskites associated with BaO and VO$_{2}$ layers.
In Section III, we briefly detail the LSDA+U electronic structure
calculation methods we have employed to model LTO/BVO/LTO or BVO/LTO/BVO quantum well systems.
The combination (LTO/BVO) was chosen because LTO is polar and BVO is not and
because this material combination should not lead to large strains at the interface.
Partly for convenience and partly because multi-quantum-well
systems are likely to be of equal experimental interest, we assume periodically repeated quantum wells so that
we have a superlattice in the growth direction.  In bulk both LTO and BVO are $d^{1}$ Mott
insulators.  We focus on heterojunction systems composed
of materials which have the same $d$-band occupancies in the bulk because more possibilities
exist for interface electronic structure compared to the case of heterojunctions between
$d^{1}$ and $d^{0}$ materials such as LaTiO$_3$/SrTiO$_3$, for which interface metallicity appears
to be inevitable.  In Section IV we present the results of our {\em ab initio} electronic structure calculations for
the two representative
quantum well systems and compare with the conjectures we made in section II.
Finally in Section V we briefly summarize our results.

\section{Polar-Nonpolar quantum well structures}

\begin{figure}
\includegraphics[width=2.50in]{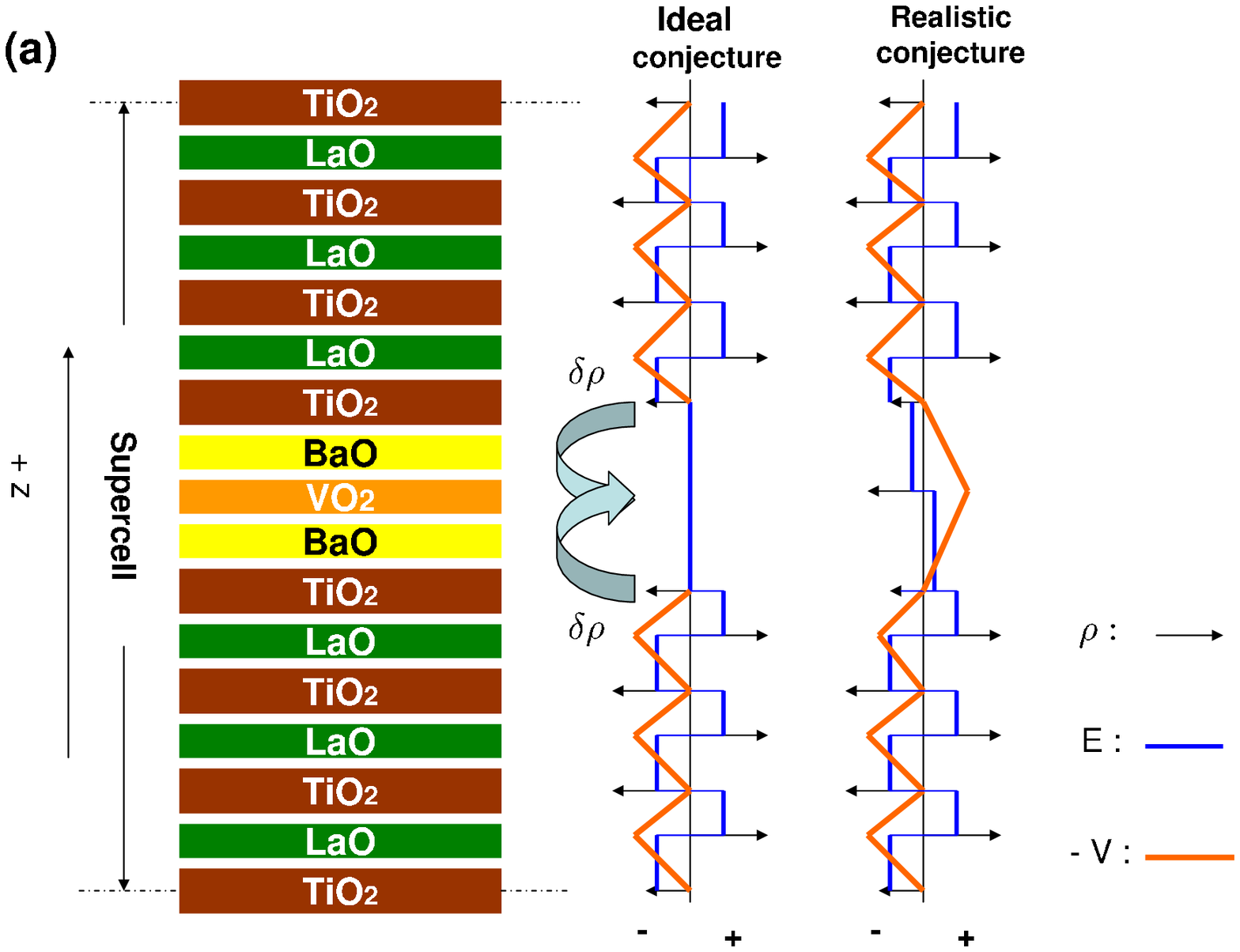}
\hfill
\includegraphics[width=2.45in]{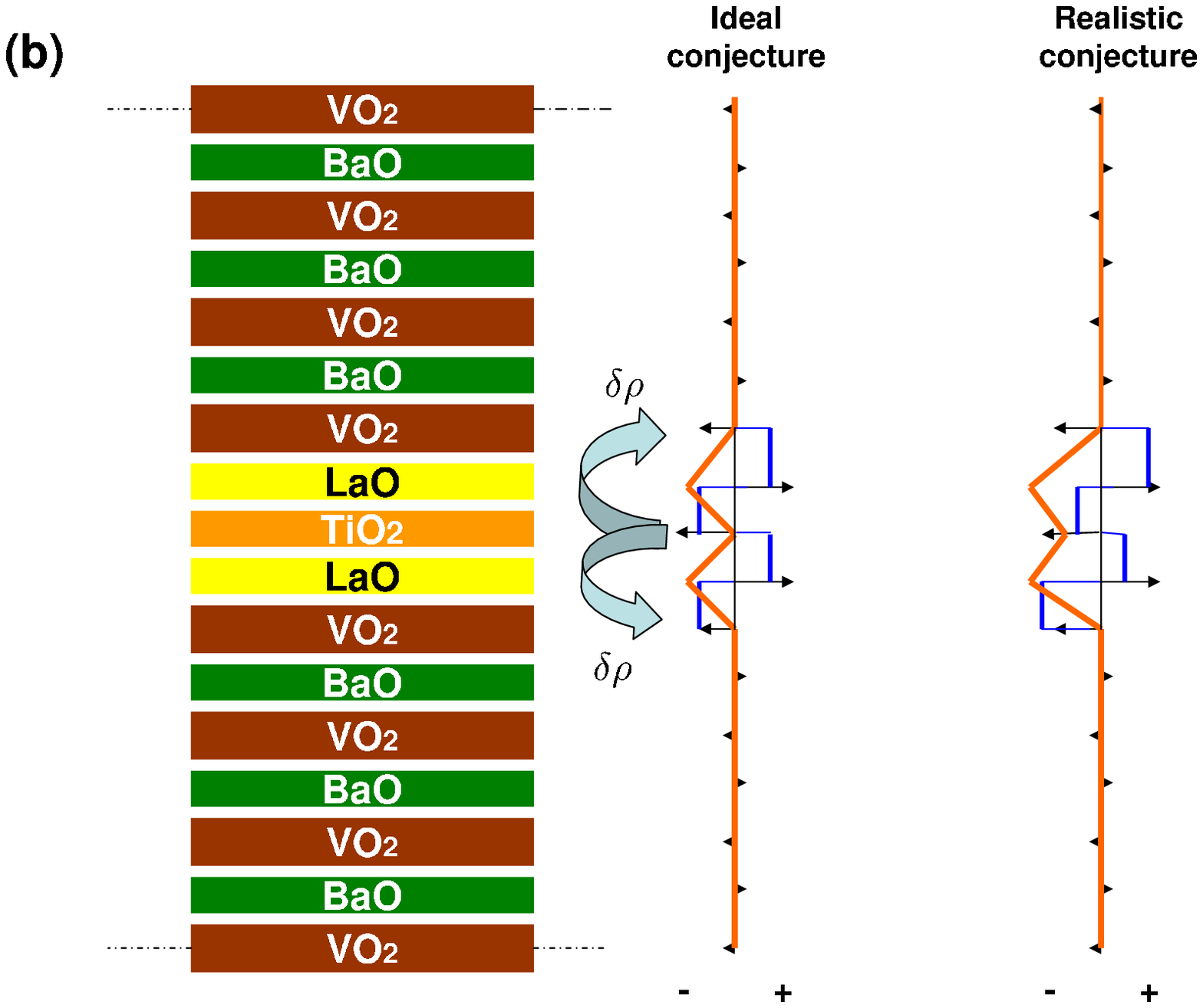}
\hfill
\includegraphics[width=2.45in]{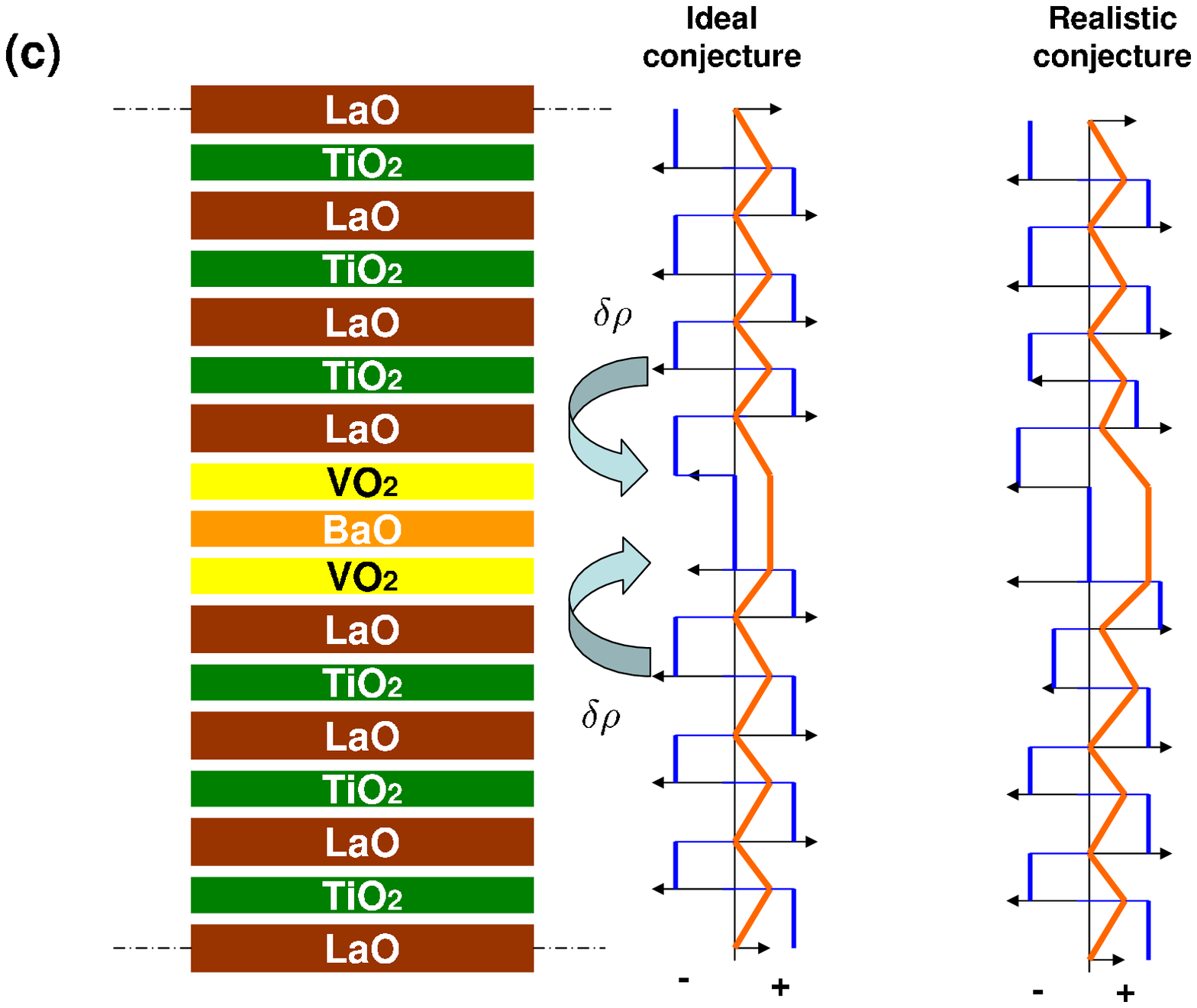}
\hfill
\includegraphics[width=2.45in]{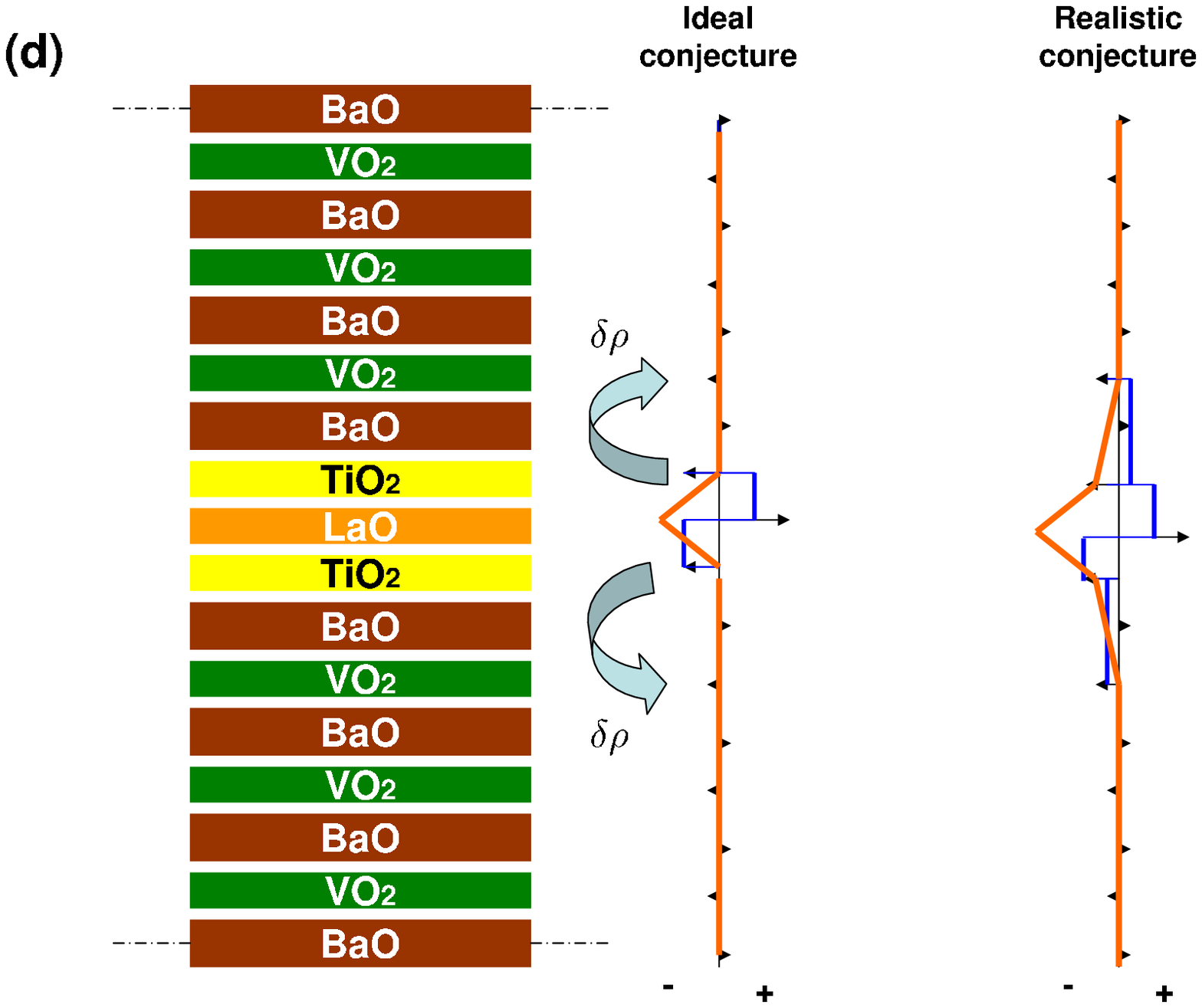}
\caption{(color online) Approximate charge density, electric field, and electric potential distributions in
different polar-nonpolar quantum well systems with inversion symmetry in the quantum well center.  The arrows (black on line)
indicate the expected charge densities $\rho$ centered on different atomic layers, the piece-wise constant thick lines lines (blue on line)
indicate the expected electric fields $E$ between the atomic layers, and the cusped lines (red on line) indicate the
expected electric potential $-V(z)$ distribution.
 The signs $+$ and $-$ stand for the direction of electric field $E$, the value for $\rho$, and the potential $-V(z)$.
 }\label{fig:Fig1}
\end{figure}

In this section, we discuss four different possible geometries for a polar-nonpolar
quantum well system.  The main points we wish to make are shown in Fig.~\ref{fig:Fig1}
which illustrates approximate charge transfers and electric potentials.  Throughout this discussion we
use LTO to represent a prototypical $d^{1}$ polar perovskite and BVO to represent a prototypical
$d^{1}$ nonpolar perovskite.  We identify two possible geometries for both
LTO/BVO/LTO (case (a) and case (c))and BVO/LTO/BVO (case (b) and case (d))quantum well systems.
At an interface between two strongly correlated bulk materials
we should expect a complex
interplay between the ionic potential energy, kinetic energy, and correlation
energies of the two materials.
By using density functional theory with local correlations (LSDA+U)
we can take into account those energies approximately with a single electron picture.
The electronic structure is characterized most crudely by an ionic electron counting picture which neglects
hybridization between oxygen and transition metal orbitals and ignores the finite thickness of the various atomic layers.
Using this picture, the La$^{+3}$O$^{-2}$ planes in LTO
are positively charged and the Ti$^{+3}$O$_{2}^{-4}$ are negatively charged.
In bulk BVO on the other hand, the Ba$^{+2}$O$^{-2}$ and V$^{+4}$O$_{2}^{-4}$ planes are neutral in the bulk.
The transition-metal ions in both materials then have one electron in the $d$ orbitals to
maintain charge neutrality.
When we combine polar perovskites with nonpolar ones,
the transition-metal layer will distribute the $d$-charges in a different way to compensate
for the nonuniform distribution of
ionized layers in this system, as illustrated in Fig.~\ref{fig:Fig1}.
The distributions labeled {\em ideal conjecture} in this
figure assign the charge deficiency or excess to the two transition metal layers adjacent to the quantum well whereas
the distributions labeled {\em realistic conjecture} assume additional charge redistribution $\delta \rho$ due to the difference
between V and Ti $d$-orbital energies.
We discuss only cases in which the center of the quantum well is a center of inversion symmetry.
In each case the number of layers which would have a positive charge in a bulk polar crystal differs from the
number of layers which would have a negative charge.

In case (a), we have two BaO layers and one VO$_{2}$ layer inserted in the
unit cell of a superlattice with six LaO$^{+}$ layers but seven TiO$_{2}$ layers.
The charges in the supercell cannot be neutral if we assume that each TiO$_{2}$ is negatively charged as in
the bulk LTO states.  Instead,
we expect that the extra electrons will be distributed evenly
among the TiO$_{2}$ layers near the central VO$_{2}$ layer.
The easiest way to accomplish this is to have the
TiO$_{2}$ layers next to BaO layers to be Ti$^{+3.5}$O$_{2}^{-4}$
rather than Ti$^{+3}$O$_{2}^{-4}$, {\em i.e.}  that they
contain 0.5 $d$-electrons per Ti rather than 1.0 $d$-electrons
per titanium.  This idealized conjecture ignores $d-p$ hybridization as well
as the difference between the $d$-orbital energies on Ti and V sites.  In particular a
V$^{+4}$ ion should have a lower intrinsic atomic levels than the Ti$^{+3}$ ion.
We should expect that charges might react locally
to this difference by transferring additional electrons $\delta \rho$ to the VO$_{2}$ well from nearby TiO$_{2}$ layers.
This additional charge transfer partially screens the original atomic energy level difference.
These considerations are illustrated schematically in Fig.~\ref{fig:Fig1}.
The $d$-electron system electronic structure reconstructs in this way near the interface in reaction to the
electrostatic potentials produced by the bulk crystal layer polarizations.
Qualitatively similar considerations would apply to wider BVO quantum wells.

In case (b), we have two LaO$^{+}$ layers and one TiO$_{2}$ barrier layer in the middle of the supercell. By the same arguments as case (a),
the simplest conjecture is that the VO$_2$ layers nearest the TiO$_2$ layers will be
negatively charged (V$^{+3.5}$O$_{2}^{-4}$) to keep the system neutral.
The  charge configuration Ti$^{+3}$O$_{2}^{-4}$ is maintained by the protection of two surrounding LaO$^{+}$ layers.
Therefore, there will be 1.5 $d$-electrons on V sites in the V$^{+3.5}$O$_{2}^{-4}$ layers and 9 $d$-electrons in the supercell.
However, the local physics is different in this case. The electron can respond locally by transferring electrons from the barrier TiO$_{2}$ to the VO$_{2}$ well layers. The reconstructed potential should therefore be lower in the well region in this case.

In case (c), we switch the ordering of layers from case (a) (BaO $\leftrightarrow$ VO$_{2}$ and LaO  $\leftrightarrow$ TiO$_{2}$).
There are 7 positive charged LaO$^{+}$ layers and 6 Ti$^{+3}$O$_{2}^{-4}$ layers. To compensate for the charge difference,
VO$_{2}$ layers have to be negatively charged as V$^{+3.5}$O$_{2}^{-4}$ in the idealized conjecture illustrated.
Additional charges can
transfer from adjacent Ti$^{+3}$O$_{2}^{-4}$ to the $V^{+3.5}$O$_{2}^{-4}$ layers.
The reconstructed potential should therefore be higher
in the well region.

Similarly in case (d), the structure has been changed from case (b) by switching the ordering
BaO $\leftrightarrow$ VO$_{2}$ and LaO $\leftrightarrow$ TiO$_{2}$. There is only one LaO$^{+}$ layer in this case.
The abutting TiO$_{2}$ layers should have the configuration Ti$^{+3.5}$O$_{2}^{-4}$ to keep charge neutrality,
resulting in a lower reconstructed local potential in the center of the cell.
Notice that the magnitude of the additional charge transfer and its
spatial distribution expected in the
realistic pictures needs to be determined by self-consistent microscopic calculations.
In the next section, we will focus on cases (a) and (b) to check the degree to which
the qualitative pictures explained here agrees with the results of microscopic DFT calculations and to
obtain further insight into the nature of the two-dimensional electron systems which
these electronic reconstructions enforce. To avoid confusion to the readers, in microscopic calculation, the $d$ electrons should be
interpreted as the electrons occupying the bands, which hybridized with $s$, $p$ orbitals, with dominant $d$ character at transition-metal sites.

\section{Electronic Structure Calculations}
In our {\em ab initio} electronic structure calculations we have considered symmetric quantum
well geometries only, use the experimental atom positions\cite{Koreans,data} for LTO,
and neglect atomic relaxation in the BVO layers. (The room temperature crystal structure of BVO is not known
experimentally.)
The electronic structure calculations were performed using DFT with LSDA+U as implemented in the software package VASP
\cite{VASP1,VASP2,VASP3}.
Projector-augmented wave pseudopotentials\cite{PAW} are used to describe the
electron-ion interaction. We sample the full superlattice Brillouin
zone (BZ) with a 5 $\times$ 5 $\times$ 3 mesh and used an energy cutoff of
410 eV which we found to be sufficient to reproduce
bulk LTO properties.
We used PAW pseudopotentials supplied by the VASP code for La, O, Ba, Ti and V but for Ba, Ti and V, we used the one with  semicore states treated as valance states.
The screened local correlation $U$ in the partially filled $d$
bands of the transition metal elements were treated using the rotationally invariant LSDA+U method due to Dudarev et al.{\cite{LSDA+U}}.
Since the La $f$ bands lie much higher in energy in experiments\cite{La, Spaldin}
we impose a much larger $U_{La}=11$ eV on La $f$ states to prevent their mixing with $d$ bands at low energy.
We chose  $U_{Ti} = 3.2$ eV and $J_{Ti} = 0.9$ eV since these values reproduces the Mott-insulating nature of bulk LTO\cite{EPL} and
we take $U_{V}$ and $J_{V}$ to be the same as $U_{Ti}$ and $J_{Ti}$. We have
varied $U_{V}$ to check the robustness of our calculations
and find that our conclusions are independent of this LSDA+U parameter.
\begin{figure}
\includegraphics[width=3in]{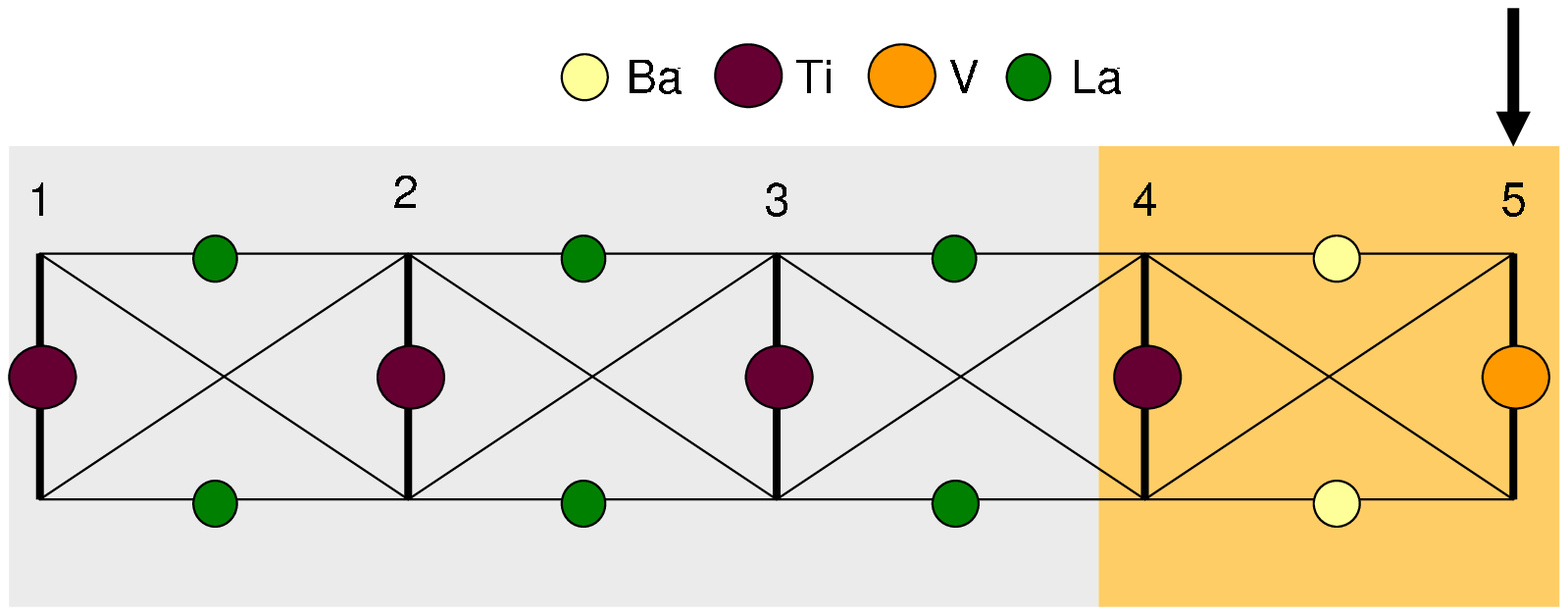}
\caption{(color online) Half of the inversion symmetric unit cell.
The quantum well consists of two layers of BaO and a monolayer of VO$_{2}$ and is
embedded in bulk LaTiO$_{3}$.  The light-orange shadowed region indicating the narrow
active region in which the electronic interface reconstruction (EIR)
driven by the polar discontinuity takes place.
Oxygen atoms are located at the intersections
of the checkered lines and form octahedra
around the Ti and V atoms. The numbers specify labels for different layers.
The downward arrow indicates the location
of the VO$_{2}$ inversion plane. }\label{fig:Fig2}
\end{figure}

\section{LSDA+U quantum-well electronic structure}
\subsection*{Case A : (LaTiO$_{3}$)$_{3}$/(BaO)$_{2}$VO$_{2}$/(LaTiO$_{3}$)$_{3}$TiO$_{2}$}
The unit cell of the superlattice consists of a single VO$_2$ layer at the center of the well surrounded by
two BaO layers as illustrated in Fig.~\ref{fig:Fig2}.  Because of the smaller $p$-$d$ energy separation in bulk
BVO compared to LTO, this material combination should form a quantum
well centered on the VO$_{2}$ layer, assuming that the $p$-bands of the two materials are nearly aligned at the
interface.  In our superlattice the BVO quantum well layers are separated
by 7 TiO$_2$ transition metal layers, each in turn separated by a LaO layer.
Half of the inversion symmetric unit cell
is shown in Fig.~\ref{fig:Fig2} in which the VO$_{2}$ inversion plane is marked by an arrow.
Ti and V in bulk LTO and BVO respectively both have nominal valence charge $d^{1}$.
In the quantum well structure we study here the number of LaO layers, which are
positively charged in the polar LTO crystal, is one smaller than the number of
TiO$_2$ layers.  It follows that the total $d$ charge per transition metal
summed over all layers is expected to be reduced by $1$.  Unless the inversion symmetry is
broken we should expect the missing $d$-charge to be symmetrically distributed around the
quantum well center.  The simplified assumptions of the previous section would
divide this valence change between the two TiO$_2$ layers adjacent to the quantum well.

\begin{figure}
\includegraphics[width=3in]{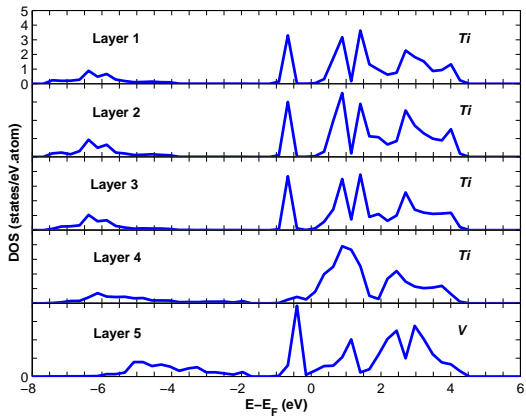}
  \caption{(color online) Layer-resolved $d$-projected density of states at Ti and V sites versus energy.
  The layer labeling is defined in Fig.~\ref{fig:Fig2}.}\label{fig:Fig3}
\end{figure}

Fig.~\ref{fig:Fig3} shows the $d$-projected DOS at a Ti site and a V site in each TiO$_{2}$ and VO$_{2}$ plane.
The states below $- 1.5$ eV are the $d-p$ bonding states between Ti, V atoms and their neighboring oxygens
which have dominant $p$-character,
whereas the states near the Fermi level (E $=$ E$_{F}$) are the  anti-bonding states which have dominant $d$-character.
As we move through the superlattice from TO$_{2}$ barrier layers to the VO$_{2}$
we see the DOS evolve from the typical bulk LTO form (layers $1$ to $3$) to
a new form near the heterojunction.  In particular,
there is a large depletion away from atomic-like $d^{1}$-charges
in layer 4.
The electronic reconstruction which accommodates the band offsets and the
polarity discontinuity is evident very strongly in the layers 5 and 6 surrounding
the quantum-well layer 4 (layer 6 is not shown in Fig.~\ref{fig:Fig3} to avoid redundancies because of the inversion symmetry of the system).
We also observe that the $p$-$d$ separation is maintained at the
VO$_2$ layer as expected after the electronic reconstruction.
The total electron transfer from layer 4 is larger than in the {\em ideal conjecture}, as
expected, and more remote layers participate in the transfer because of oxygen mediated coherence between transition
metal orbitals centered on different layers.

\begin{figure}
\includegraphics[width=3in]{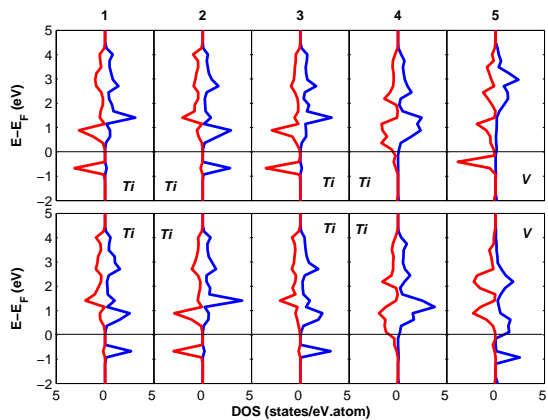}
  \caption{(color online) Spin and layer resolved densities-of-states versus energy. The labels on top are layer labels as in Fig.~\ref{fig:Fig2}.
  Since we allow G-type AFM order in the barrier material, there are two distinct transition metal atoms in each layer.
  The DOS for majority spins and minority spins are represented by
  red and blue lines respectively and the Fermi level is marked with a horizontal dark line at $E=E_{F}$.}\label{fig:Fig4}
\end{figure}

Most Mott insulators are orbitally and magnetically ordered.  In order
to gain some insight into the way in which the magnetic order is disturbed by the
quantum well structure, we plot in Fig.~\ref{fig:Fig4} the spin-resolved $d$-projected DOS in
all layers.  In our calculations we find G-type antiferromagnetic (AFM) order in which neighboring transition metals
ions have opposite spins both within and between layers for the bulk like TiO$_{2}$ layers. (layers 1, 2, 3 in Fig.~\ref{fig:Fig4}).
Layer 4 on the other hand has an overall magnetic
moment of around 0.1 $\mu_{B}$ per Ti ion indicating a ferromagnetic layer.
Because of the electronic interface reconstruction (EIR), the magnetic order in the VO$_{2}$ plane changes from AFM to ferromagnetic.
The projected moments on each Ti ions in the AFM bulk-like TiO$_{2}$ layers
is calculated to be 0.74 $\mu_{B}$ inside the Ti ionic radius  which is fixed in our calculation by the
pseudopotential construction. We believe that by enlarging the Ti ionic radius, we can get $d$-orbital integrated moments
close to the 1 $\mu_{B}$ value expected for $d^{1}$ Mott insulators.
Note that ferromagnetism is induced in the VO$_2$ layer by the two adjacent strongly depopulated and
ferromagnetic TiO$_2$ layers.

\begin{figure}
\includegraphics[width=4.5in]{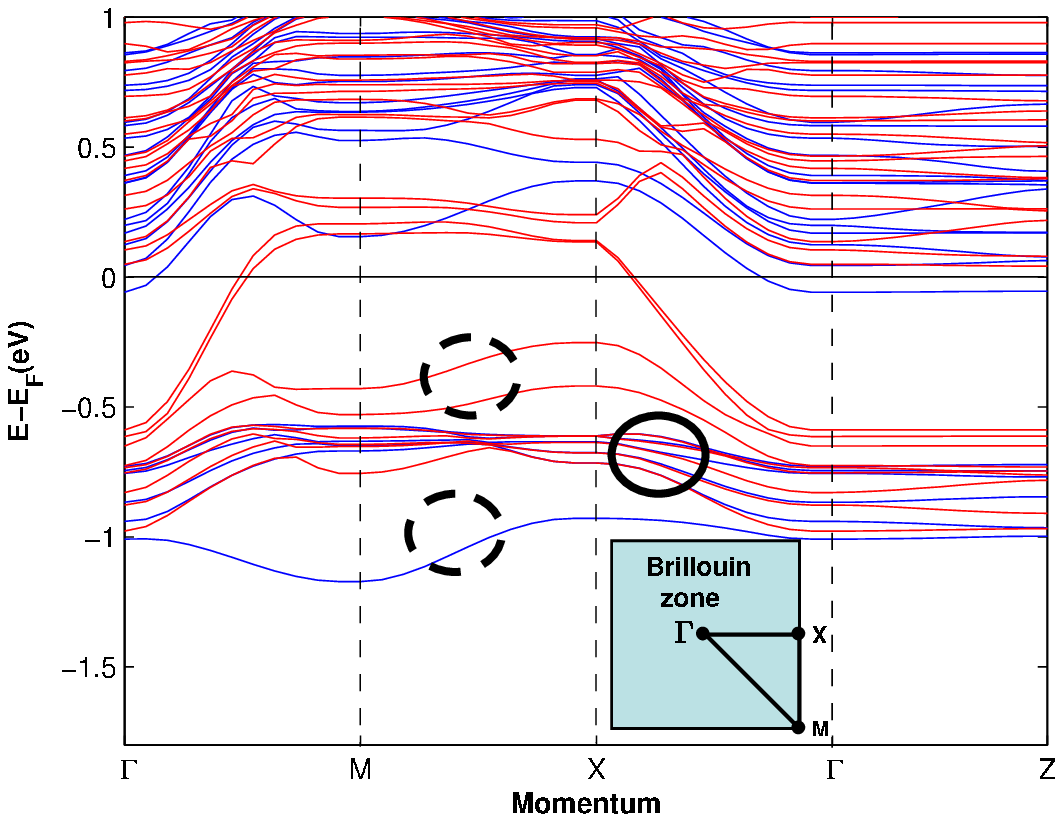}
  \caption{(color online) Spin-resolved band structure at the Brillouin zone. The horizontal axis represents superlattice crystal momentum.
  The red solid lines represents the majority-spin bands and the blue ones are the minority-spin bands.
  The momentum dependence along the layer direction $k_{z}$ is indicated in the region between the $\Gamma$ and Z points.
  The full circle marks the $d^{1}$ bands for Ti atoms in LTO bulk states. The dashed circles represents the $d$ bands for V atoms in VO$_{2}$.
The bands associated with EIR are flat along k$_{z}$ demonstrating their quasi-two-dimensional character.
The shadowed blue square represents the projection of the superlattice Brillouin zone in the plane of the quantum well.
}\label{fig:Fig5}
\end{figure}

Fig.~\ref{fig:Fig5} shows the superlattice band structure along high symmetry directions in the Brillouin zone.
The bands (five majority (red) spin bands and five minority (blue) spin bands), which
lie far below the E$_{F}$ (in the energy range between
between $- 1$ eV and $- 0.6$ eV), can be identified as bulk LTO $d$-charge character
bands by comparing with Fig.~\ref{fig:Fig4} and
calculating band-decomposed local charge density contributions. In the supercell, we have 5 bulk TiO$_{2}$ layers (layer 1, 2, 3, 7, 8)
in our superlattice corresponding in the antiferromagnetic state to 10 full bands (5 spin-up and 5 spin-down).
Therefore, there should be ten full bulk like LTO $d$-bands, identified
by the solid circle, in Fig.~\ref{fig:Fig5}.
Two full majority spin bands and one full minority spin band (near -1 eV) highlighted by two dashed circles are associated
with the $d$-charges in the VO$_2$ layers.
The two majority bands close to the interface states in energy and are hybridized along the layer-growth direction and delocalized
(This interpretation has been verified by separating the charge density contribution of those bands) .
Because of this hybridization, the total $d$-charges (2.4053 $e^{-}$) at the V sites is smaller
than the 3 $e^{-}$ value which would be implied by
a local interpretation of the band structure.
It can in general be misleading to estimate the number of charges at transition metal ions by counting the number
of reconstructed bands with a particular dominant orbital character.
The three partially full bands (one minority band and two majority bands) near the fermi surface E$_{F}$ are the bands
responsible for the charge depletion at the TiO$_{2}$ layers adjacent to the
quantum well (layer 4 and 6).
The momentum k$_{z}$ dependence of the electronic states is shown in the right panel of Fig.~\ref{fig:Fig5} (from momentum (0 , 0, 0) at $\Gamma$ to (0, 0, $k_z$) at Z). It shows that, the states at the Fermi energy form a two-dimensional electron gas
centered on three layers, the active region.
Notice that there also appears a flat band between M and X, which corresponds to the direction away from the bonding direction between Ti and O atoms
in charge depleting the TiO$_{2}$ layers, due to the fact there is negligible tunneling
of electrons along this direction.
The two dimensional system is metallic in this case and has three partially filled bands.

\begin{figure}
\includegraphics[width=3in]{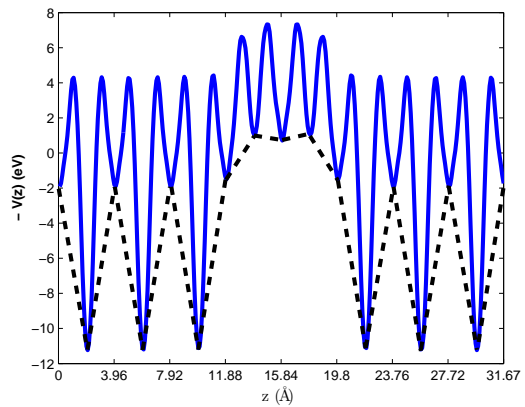}
  \caption{(color online) Plane averaged Hartree potential along the layer-growth direction $z$.
  The blue curve represents the Hartree potential experienced by electrons $-e V(z)$.
  The dashed line indicates the net potential variation between charge-concentrations centers
in each layers.}\label{fig:Fig6}
  \end{figure}

Since we have discussed the local potential for the electron in case (a) in Fig.~\ref{fig:Fig1} based on qualitative model
in which we assume the charges at each layer are concentrated in a single layer along the layer direction, we would like to examine how
the the crude model prediction compares with actual DFT predictions.
In Fig.~\ref{fig:Fig6}, the DFT plane-averaged Hartree potential for the electrons is shown.
First TiO$_{2}$ layer starts at $z = 0$ and layers repeat with a spatial period of 3.96 ${\AA}$ and with the
central layer ($z = 15.84$ ${\AA}$) of the supercell substituted by the VO$_{2}$ layer.
First LaO layer is located at $z = 1.98$ $\AA$ and these layers repeat
with a period of 3.96 ${\AA}$ except that the two layers surrounding the central VO$_{2}$ layer
correspond instead to BaO layers.
By inspecting the variation of the local minima in the potential, we can see that the qualitative behavior
anticipated in case (a) in Fig.~\ref{fig:Fig1} is valid, as shown by the black-dashed line. The
electric field in the qualitative model should be considered as a macroscopically averaged electric
field between potential minima in the microscopic calculations.
Notice that the potential of the VO$_{2}$ layer is somewhat lower than anticipated.

\begin{figure}
\includegraphics[width=3in]{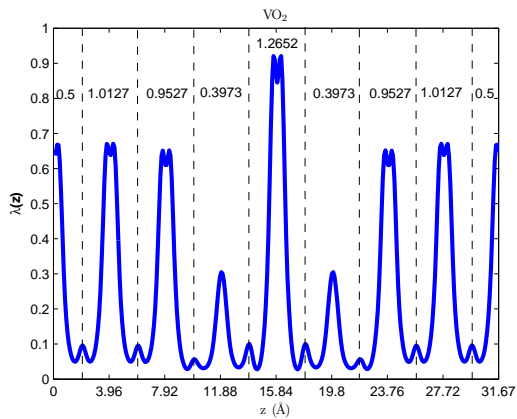}
\caption{(color online) Linear electron number density $\lambda(z)$ along layer-growth direction z. The dashed lines represent
the boundaries used for the partitioning of density into contributions from different layers.
The total number of electrons in each layer contributed by the $d$-character bands in Fig.~\ref{fig:Fig5} is
indicated by the numerical values
between dashed lines. In this case, the total number of electrons in the supercell is seven.}\label{fig:Fig7}
\end{figure}

In Fig.~\ref{fig:Fig7}, we show the linear electron number density $\lambda(z)\equiv 0.5 \int \rho(x,y,z) dx dy$ along the layer-growth direction.
The charge number spatial distribution $\rho(x,y,z)$ is normalized per transverse unit cell which is doubled because of the system's
antiferromagnetic order, and includes the total number of electrons contributed by the $d$-bands between
with energies up to $1.5$ eV below the Fermi level.  (See Fig.~\ref{fig:Fig5}.)
The factor $0.5$ is used to convert the normalization to $d$-electrons per transition metal ion,
to facilitate comparison with the qualitative picture.
The dashed lines represent the boundaries for the transition-metal oxide layers used in this
electron number construction.   The area under the curve $\lambda(z)$  represent the
total number of electrons from all $d$-bands.
We see an apparent charge depletion at the TiO$_{2}$ layers centered at $z = 11.88$ ${\AA}$ and $z = 19.8$ ${\AA}$ along with large charge accumulation at the VO$_{2}$ layer. There is only minor charge depletion at second-nearest TiO$_{2}$ layers ($z = 7.92$ ${\AA}$ and  $z = 23.76$ ${\AA}$) away from VO$_{2}$ layer.
The total charges from the $d$-bands, 7 $e^{-}$, is obtained by adding the number of electrons at each layers
and agrees with the qualitative predictions for case (a).
The number of electrons in the depleted TiO$_{2}$ layers is $0.3793$. The other layers have small deviation from one electron.
The calculation shows a clear deviation from ideal conjecture we made for this case and is closer
to the {\em realistic conjecture}.
In general, the charges in the well are mainly transferred from the adjacent TiO$_{2}$ layers but there is also
a minor contribution from
other TiO$_{2}$ layers as shown from the number of electrons in each layers.

\subsection*{Case B : (BaVO$_{3}$)$_{3}$/(LaO)$_{2}$TiO$_{2}$/(BaVO$_{3}$)$_{3}$VO$_{2}$}
In this section, we consider the quantum well structure shown in case (b). In this case, the number of LaO layers, which are
positively charged in the polar LTO crystal, is greater than the number of
TiO$_2$ layer by one.  It follows that the total $d$-electrons per transition-metal
summed over all layers is expected to be increased by $1$.  Unless the inversion symmetry is
broken we should expect the extra $d$-charge to be symmetrically distributed around the
quantum well center.

\begin{figure}
\includegraphics[width=3in]{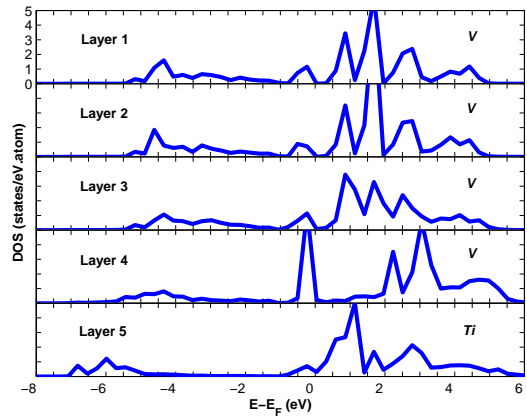}
\caption{(color online) Layer-resolved $d$-projected density of states at Ti and V sites versus energy.}
\label{fig:Fig8}
\end{figure}

\begin{figure}
\includegraphics[width=3in]{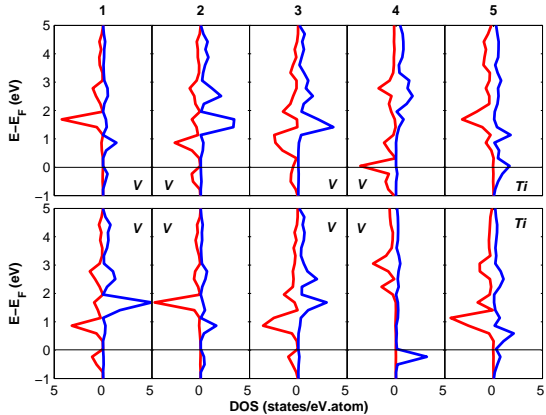}
  \caption{(color online) Spin and layer resolved densities-of-states versus energy. The labels on top are the layer labels
  indicated in Fig.~\ref{fig:Fig2}.
  Since we allow G-type AFM order in the barrier material, there are two distinct transition-metal ions in each layer.
  The DOS for majority spins and minority spins are represented by
  red and blue lines respectively.}\label{fig:Fig9}
\end{figure}

\begin{figure}
\includegraphics[width=3.5in]{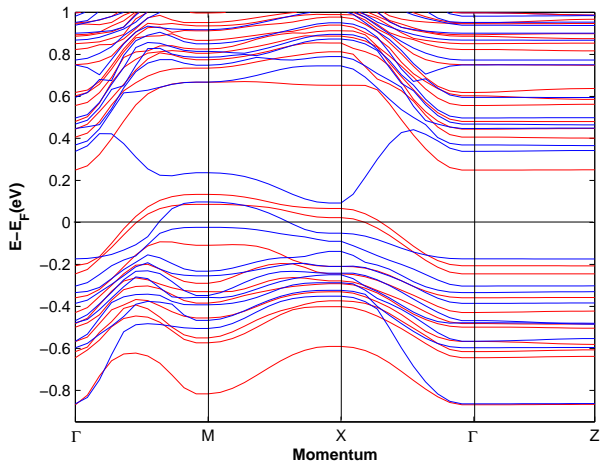}
  \caption{(color online) Spin-resolved band structure. The horizontal axis represents superlattice crystal momentum.
  The red solid lines represent the majority-spin bands while the blue lines represent minority-spin bands.
  The momentum dependence along the layer direction $k_{z}$ is captured in the region between the $\Gamma$ and Z points.
  The three partially occupied bands associated with EIR are flat along k$_{z}$, demonstrating
  their quasi-two-dimensional character.}\label{fig:Fig10}
\end{figure}

Fig.~\ref{fig:Fig8} shows the $d$-projected DOS at a Ti site and a V site in each TO$_{2}$ and VO$_{2}$ plane.
The states below $-1$ eV are the $p-d$-bonding states between the transition-metal atoms (Ti, V) and the $p$ orbitals of the neighboring oxygens
and have dominant $p$ character.
The states near the Fermi level (E $=$ E$_{F}$), on the other hand, have dominant $d$-character.
As we move through the superlattice from the VO$_{2}$ layers to the TiO$_{2}$ barrier layer
we see the DOS evolve from the typical bulk BVO layer form (layers $1$ to $3$) to
a new form near the heterojunction.  In particular, we see a strong peak in DOS
in layer 4 near the Fermi energy $E_F$ with 1.6 $e^{-}$ per V ion in agreement with our conjectures.
As far as magnetic order is concerned (Fig.~\ref{fig:Fig9}),
we find G-type AFM order in which neighboring transition metals
ions have opposite spins both within and between layers for the bulk like VO$_{2}$ layers (layers 1 and 2).
The magnetic orders in layer 3 and 5 are reconstructed to be ferromagnetic with the magnetic moment 0.9755  $\mu_B$ per V ion and the moment 0.4750  $\mu_B$ per Ti respectively.
The magnetic order in layer 4 is ferrimagnetic with a net magnetic moment 0.2170 $\mu_{B}$.  In
the corresponding band structure illustrated in Fig.~\ref{fig:Fig10}, we
identify three partially filled
two-dimensional bands that cross
the Fermi level, and sixteen full $d$-bands.
These bands accommodate 9 $d$-electrons obtained by taking the average $d$-charges per transition metal ion and summing the charges of all the layers,
as illustrated in Fig.~\ref{fig:Fig12} discussed below.
The two majority spin partially-filled bands, and the single minority-spin partially-filled band, are
responsible for the extra charge at V sites in layer 4 and 6 as shown in Fig.~\ref{fig:Fig9}.
The three full minority-spin bands, which are lower in energy than the partially-filled minority-spin band
but closer to the Fermi level than other full minority-spin bands
mainly contribute to minority-spin charge concentrated in layers 4, 5, and 6.
The wave functions for those bands are
strongly hybridized within those layers.
The lowest majority-spin band in Fig.~\ref{fig:Fig10}
is mainly of V character in layer 4 but lower in energy due to the
more attractive local potential at this site.
The peak of the DOS for the minority spin at the V site seems to shift higher toward
the Fermi energy than expected.
This is due to the fact those states are strongly hybridized with the states at the
Ti sites which are shifted toward higher energy, an effect not included in the qualitative analysis.
 (See the potential profile of the realistic conjecture of case (b) in Fig.~\ref{fig:Fig1}).
Other full bands are responsible for BVO bulk states away from the region in which
the electronic interface reconstruction occurs.

\begin{figure}
\includegraphics[width=3in]{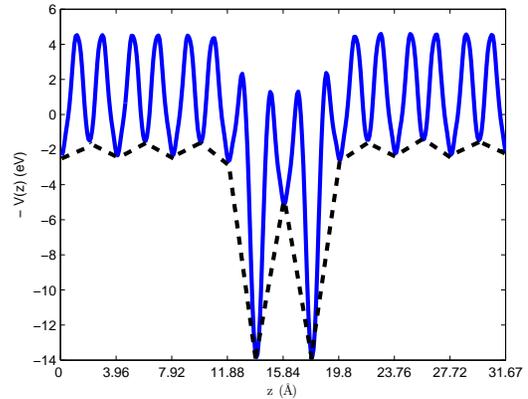}
  \caption{(color online) Plane averaged Hartree potential along the layer-growth direction $z$.
  The solid dashed line indicates the net potential variation between charge-concentrated centers at each layer.}\label{fig:Fig11}
\end{figure}

\begin{figure}
\includegraphics[width=3in]{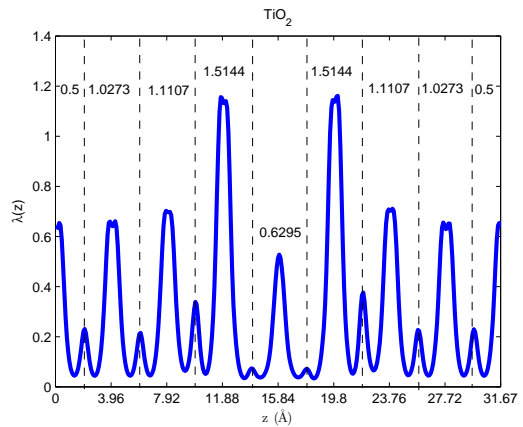}
\caption{(color online) Linear electron number density $\lambda(z)$ along layer-growth direction z. The dashed lines represent
the boundaries between different layers used in constructing the layer decomposition of this charge density.
The numerical numbers
between dashed lines in each layers are the total number of electrons from $d$-derivative bands in Fig.~\ref{fig:Fig10}.
In this case, the total number of electrons is nine.}\label{fig:Fig12}
\end{figure}

In Fig.~\ref{fig:Fig11}, we plot plane-averaged Hartree potentials.
The first layer starts with VO$_{2}$ at $z = 0$.  Transition metal layers repeat with a spatial period of 3.96 ${\AA}$
with the center ($z = 15.84$ ${\AA}$) layer occupied by TiO$_{2}$ layer.
At $z = 1.98$ $\AA$, the first BaO layer starts and repeats also at a period of 3.96 ${\AA}$; the local minima closest to TiO$_{2}$
correspond instead to LaO layers.
By inspecting the local minima in the Hartree potential, we can see that the
DFT calculations largely verify the scenario of case (b) in Fig.~\ref{fig:Fig1} except that the dashed line wiggles in the region of bulk BVO. The wiggling may due to the slight difference of the strength of ionic bonding between Ba-O and V-O along
the layer direction, which causes the small dipole modification of charge distribution in BVO ionized layers.

In Fig.~\ref{fig:Fig12}, we show the corresponding electron linear number density $\lambda(z)\equiv 0.5 \int \rho(x,y,z) dx dy$ along the layer-growth direction.
This charge number spatial distribution $\rho(x,y,z)$ include the total charges contributed from the bands between 0 and - 0.9 eV with respect to Fermi level in Fig.~\ref{fig:Fig10}.
We see an apparent charge accumulation (1.5144 $e^{-}$) at the TiO$_{2}$ layers centered at $z = 11.88$ ${\AA}$ and $z = 19.8$ ${\AA}$ along with much less charge (0.6295 $e^{-}$) at the TiO$_{2}$ layer.
Remote transition metal layers have small deviations from a $d^{1}$ valence.
The total amount of charges from the $d$-bands is 9 $e^{-}$ by adding the charges at each layer in agreement with our prediction in case (b).
The calculation shows small deviations from the ideal conjecture we made for this case and almost the realistic conjecture.
In general, the charges in the well are transferred from the TiO$_{2}$ layer to adjacent VO$_{2}$ layers as discussed in section II.

\section{Summary and Discussion}
In this paper we have explored some possibilities
for the realization of artificial two-dimensional electron systems
by growing quantum wells consisting of a few atomic layers of one layered
oxide surrounded by another material.  In particular we consider systems
which involve heterojunctions between polar insulating perovskites like LTO and
nonpolar insulating perovskites like BVO.  Perovskites can be viewed as consisting of
alternating AO and BO$_2$ layers where B is a transition metal.
The main idea of this paper is that inversion symmetric
quantum wells will necessarily have a different number of
AO and BO$_2$ layers, and that this will necessarily lead to
an electronic reconstruction which induces
two-dimensional electron systems near the quantum well when
one material is polar and one is nonpolar.
Four different cases can be identified, in which the quantum well
material is either polar or non-polar and in which it either has an
excess AO layer or an excess BO$_2$ layer.
The excess layers induce either a unit increase or a unit decrease in the total
$d$-orbital occupancy per transition metal site when summed over
all layers.  This charge must be localized close to the quantum well to avoid
large band shifts due to polarization driven electrostatic potentials.
As long as inversion symmetry around the quantum well center
is not-broken spontaneously, half of this charge must appear on each half
of the quantum well center.  Because of the strong-correlation character of
the oxide electronic structure, these charges will tend to be localized
in half-filled LSDA+U bands which are weakly hybridized so that both cross the
Fermi energy.

We have tested and elaborated on this qualitative picture of the electronic structure of polar-nonpolar heterojunction
quantum wells in perovskites by performing self-consistent LSDA+U electronic structure calculations for
a representative system composed of BVO and LTO without any atomic relaxation at the interface.
The LSDA+U calculations capture the interplay between long range polar electrostatic energy and local physics driven by
energetic offsets between different transition metal ions and by $p-d$ hybridization.
We find that in the two cases we have examined one additional two-dimensional bands is induced by the energetic offsets.
In all cases the quantum well structure creates a two-dimensional metal inside a three dimensional Mott insulator,
and alters the character of the magnetic order.
The reconstructed magnetic order is ferromagnetic in the active well region and antiferromagnetic in bulk region.
The appearance of two-dimensional ferromagnetism in intimate exchange contact with antiferromagnetism
is intriguing from the point of view of spintronics.

%{\bf Allan:  Check this carefully.  All OK?}
Because of the complexity of oxide materials, our calculation should be viewed as an exploration of
possibilities rather than as predictive in any detailed sense.  Among the many sources of uncertainty,
the Hubbard U interaction used in our LSDA+U is a phenomenological parameter for which we have
chosen a standard bulk values.   This value should likely be renormalized near the
quantum well.  More challenging to account for is the role of inevitable lattice distortions,
oxygen stoichiometry variations, and other defects near the
interfaces.  The interfaces between the thin film oxide and the substrate on which it is grown
can also play a complicating role if either the barrier material or the substrate is polar.
For a few cases we have explored the influence of
lattice distortion effects by allowing the atoms to relax along the layer growth direction in the active region,
maintaining $U_{V}$=$U_{Ti}$$=3.2$ eV as before.
What do we find is that the amount of charge transfer to or from particular layers and
the details of the magnetic orders are highly sensitive to ionic relaxation.
These complications make the accurate prediction of the electronic reconstruction and the
associated magnetic order highly nontrivial. It is also desirable to understand if exotic symmetry breaking phases
such as superconductivity may appear in the interface .
Further progress in understanding
detailed properties of the two-dimensional electron systems we envision which almost certainly
require inputs from experiments.  The microscopic calculations reported on here can be used as guidance
in constructing phenomenological models which can be fit to any data which might emerge from future work and serve
as the theoretical basis for possible exotic phases in the interface.
The present paper has addressed $d^{1}$ transition metal systems.
Considerations similar to those explained here also apply to
transition-metal oxides with heavy transition metal elements (Cu, Ni) in which the $t_{2g}$ bands are fully
occupied far below the Fermi level, and low-energy excitations
are expressed within the $e_{g}$ bands.

\section*{Acknowledgements}
This work was supported by the National Science Foundation under grant
DMR-0606489, by the Welch Foundation
(Houston, TX) under Grants No. F-1473 and No. F-0934,
by the Texas Advanced Computing Center (TACC), University
of Texas at Austin.  B.S. thanks SRC-NRI
(SWAN) for financial support.
C.-C Joseph Wang gratefully
acknowledge helpful conversations with Peter Abbamonte.
*Electronic address: joseph@physics.utexas.edu


\begin{thebibliography}{99}
\bibitem{HS} Patrick A. Lee, Naoto Nagaosa, and Xiao-Gang Wen,
Rev. Mod. Phys. {\bf 78}, 17 (2006).
\bibitem{CM} A. P. Ramirez, J. Phys.: Condens. Matter {\bf 9}, 8171-8199 (1997).
\bibitem{MIT} M. Imada, A. Fujimori, and Y. Tokura, Rev. Mod. Phys.
{\bf 70}, 1039 (1998).
\bibitem{Huang}
A. Ohtomo, D.A. Muller, J.L. Grazul and H.Y. Hwang,
Nature {\bf 419}, 6905 (2002);
A. Ohtomo and H.Y. Hwang, Nature {\bf 427}, 423 (2004);
H.Y. Hwang, MRS Bulliten {\bf 31}, 28 (2006).
\bibitem{sc}
N. Reyren , S. Thiel , A. D. Caviglia , L. Fitting Kourkoutis
, G. Hammerl , C. Richter , C. W. Schneider ,
 T. Kopp, A.-S. Retschi  , D. Jaccard , M. Gabay ,
D. A. Muller , J.-M. Triscone  and J. Mannhart , Science
{\bf 317}, 1196 (2007).
\bibitem{Brinkman}
A. Brinkman , M. Huijben , M. Van Zalk , J. Huijben ,
U. Zeitler , J. C. Maan , W. G. Van der Wiel , G. Rijnders
, D. H. A. Blank  and H. Hilgenkamp , Nature Mat.
{\bf 6}, 493 (2007).
\bibitem{Ahn}
C. H. Ahn, A. Bhattacharya, M. Di Ventra, J. N. Eckstein, C. Daniel Frisbie,
M. E. Gershenson, A. M. Goldman, I. H. Inoue, J. Mannhart, Andrew J. Millis,
Alberto F. Morpurgo and Douglas Natelson, Rev. Mod. Phys. {\bf 78}, 1186 (2006).
\bibitem{Spaldin} Satoshi Okamoto, Andrew J. Millis, and Nicola A. Spaldin, Phys. Rev. Lett. 97. 056802 (2006).
\bibitem{Pickett}
Rossitza Pentcheva and Warren E. Pickett, Phy. Rev. Lett. {\bf 99}, 016802 (2007).
\bibitem{WC}
W.-C. Lee, A.H. MacDonald, Phys. Rev. B 74, 075106 (2006).
; Phys. Rev. B {\bf 76}, 075339 (2007).
\bibitem{Sawatzky} S. Altieri, L.H. Tjeng, F.C. Voogt, T. Hibma and G.A.
Sawatzky, Phys. Rev. B {\bf 59} R2517 (1999).; J. van den Brink
and G.A. Sawatzky, Europhys. Lett. {\bf 50} 447 (2000).
\bibitem{Koreans} Hyo-Shin Ahn, Do Duc Guong, Jaichan Lee, and Seungwu Han,
J. Korean Phy. Soc. {\bf} 49, 1536 (2006).
\bibitem{data} M. Cwik, T. Lorenz, J. Baier, R. M$\ddot{u}$ller, G. Andr$\acute{e}$, F. Bour$\acute{e}$e, F. Lichtenberg,
A. Freimuth, R. Schmitz, E. M$\ddot{u}$ller-Hartmann and M. Braden, Phys. Rev. B {\bf 68},
060401(R) (2003).
\bibitem{VASP1}
G. Kresse, and J. Furthmüller, Phys. Rev. B {\bf 54}, 11169 (1996).
\bibitem{VASP2}
G. Kresse, and J. Furthmüller, Comput. Mater. Sci. {\bf 6}, 15 (1996).
\bibitem{VASP3}
G. Kresse, and D. Joubert, Phys. Rev. B {\bf 59}, 1758 (1999).
\bibitem{PAW}
G. Kresse, and J. Hafner, Phys. Rev. B {\bf 47}, 558 (1993).
\bibitem{LSDA+U}
S. L. Dudarev, G. A. Botton, S. Y. Savrasov, C. J. Humphreys and A. P. Sutton, Phys. Rev. B {\bf 57}, 1505 (1998).
\bibitem{EPL} S. Okatov, A. Poteryaev, and A. Lichtenstein, Europhys. Lett. {\bf 70}, 499 (2005).
\bibitem{La}
M. T. Czyzyk and G.A. Sawatzky, Phys. Rev. B {\bf 49}, 14211 (1994).
\bibitem{Huang2}
N. Nakagawa, H. Y. Hwang, and D. A. Muller, Nature mat. {\bf 5}, 204 (2006).
\end{thebibliography}
\end{document}